# GHOST: A Graph Neural Network Accelerator using Silicon Photonics

SALMA AFIFI

Colorado State University, Fort Collins, CO, USA, Salma.Afifi@colostate.edu

FEBIN SUNNY

Colorado State University, Fort Collins, CO, USA, Febin.Sunny@colostate.edu

AMIN SHAFIEE

Colorado State University, Fort Collins, CO, USA, Amin.Shafiee@colostate.edu

MAHDI NIKDAST

Colorado State University, Fort Collins, CO, USA, Mahdi.Nikdast@colostate.edu

SUDEEP PASRICHA

Colorado State University, Fort Collins, CO, USA, Sudeep@colostate.edu

Graph neural networks (GNNs) have emerged as a powerful approach for modelling and learning from graph-structured data. Multiple fields have since benefitted enormously from the capabilities of GNNs, such as recommendation systems, social network analysis, drug discovery, and robotics. However, accelerating and efficiently processing GNNs require a unique approach that goes beyond conventional artificial neural network accelerators, due to the substantial computational and memory requirements of GNNs. The slowdown of scaling in CMOS platforms also motivates a search for alternative implementation substrates. In this paper, we present *GHOST*, the first silicon-photonic hardware accelerator for GNNs. *GHOST* efficiently alleviates the costs associated with both vertex-centric and edge-centric operations. It implements separately the three main stages involved in running GNNs in the optical domain, allowing it to be used for the inference of various widely used GNN models and architectures, such as graph convolution networks and graph attention networks. Our simulation studies indicate that *GHOST* exhibits at least 10.2× better throughput and 3.8× better energy efficiency when compared to GPU, TPU, CPU and multiple state-of-the-art GNN hardware accelerators.

## 1 INTRODUCTION

Deep learning has become a vital pillar in our lives due to its ability to solve many complex problems efficiently across diverse fields, including autonomous transportation, healthcare, industrial automation, and network security. This success of deep learning owes tremendously to the evolution of neural networks variants that are tailored for specific learning tasks. For example, Convolution Neural Networks (CNNs) [1] and Recurrent Neural Networks (RNNs) [2] have proven their efficiency in pattern recognition for images and sequence data, by extracting knowledge from the spatial and temporal dimensions of data. While these examples are prominent solutions for many tasks, they are limited in scope to non-arbitrary structured and Euclidean data. Arbitrary structured data, including graphs, require different techniques for efficient processing. Graph data processing is critical for many problems, e.g., social network analysis, recommender systems, and drug discovery [3].

Graph Neural Networks (GNNs) have emerged in recent years and established their proficiency in dealing with graph-structured data. These models can extract information from the graph structure and discover patterns in the data that may be difficult to identify with other deep learning methods [4]. Accordingly, many applications now benefit greatly from GNNs, and hence a lot of recent efforts have focused on enhancing GNN algorithms and improving their efficiency in handling large and various graphs. The continuing progress of GNN algorithms and models necessitates hardware platforms capable of providing GNN-specific support with high performance, while abiding by strict power constraints. Although hardware acceleration for neural networks such as CNNs and RNNs have been extensively studied, the processing of GNNs presents unique challenges due to their combination of dense and vastly sparse operations, diversity of input graphs, and the various types of GNN algorithms and models [5]. Thus, hardware accelerators tailored to accelerate the processing of conventional neural networks cannot be directly and efficiently applied to GNNs.

Moreover, relying on traditional electronic accelerators creates limitations as these platforms face challenges in the post-Moore era due to high costs and diminishing performance improvements with semiconductor-technology scaling. Moving data through metallic wires is a well-known bottleneck in these accelerators, as it restricts the achievable performance in terms of bandwidth, latency, and energy efficiency [6]. Silicon photonics technology provides a promising solution to this data-movement bottleneck, offering ultra-high bandwidth, low-latency, and energy-efficient communication [7]. Optical interconnects, which are now being considered for chip-scale integration, have already replaced metallic ones for light-speed data transmission at almost every level of computing. It is also possible to use optical components for computations, such as matrix-vector multiplication [8]. The emergence of chip-scale optical communication and computation has thus made it possible to design photonic integrated circuits that offer low-latency and energy-efficient optical-domain data transport and computation. Furthermore, prior work, from both academia and industry, has demonstrated the significant benefits resulting from using silicon photonics for the acceleration of neural networks, as in [8]-[10].

In this paper, we introduce *GHOST*, the first silicon-photonic-based GNN accelerator that can accelerate inference of diverse GNN models and graphs. The key contributions in this paper are:

- The design of a novel GNN accelerator hardware architecture using silicon photonics with the ability to accelerate multiple existing variants of GNN models;
- Detailed photonic device and circuit-level optimizations to mitigate crosstalk noise so that error-free GNN operations can be ensured in the accelerator;
- A detailed architectural optimization for the efficient handling and acceleration of diverse graph structures and GNN model architectures on the proposed hardware accelerator; and
- A comprehensive comparison with GPU, TPU, CPU, and state-of-the-art GNN accelerators.

The rest of the paper is organized as follows. Section 2 provides a background on GNNs (different models, their acceleration challenges, and previous efforts on GNN acceleration) and on silicon photonics and performing optical computations. Section 3 describes the *GHOST* architecture and our optimization efforts at the device, circuit and architecture layers. Details of the experiments conducted, simulation setup, and results are presented in Section 4. Lastly, Section 5 presents concluding remarks.

## 2 BACKGROUND

### 2.1 Graph neural networks

Prior to the emergence of GNNs, graph processing was mostly limited to traditional machine learning and graph algorithms. However, these methods had limitations in capturing the non-linear and complex relationships between vertices in a graph [4]. With the advent of GNNs, graph processing has been revolutionized, and there has been a significant improvement in graph-based machine learning tasks, such as node classification, link prediction, and graph classification. GNNs are a type of deep learning algorithm that can learn complex graph structures and relationships, and have now broadened the scope of Artificial Neural Networks (ANNs) to encompass non-Euclidean and irregular data found in graphs [5].

GNNs exploit the connections within a graph to understand and represent the relationships between vertices. They utilize an iterative approach that relies on a graph's structure and take in edge, vertex, and graph feature vectors that represent the known attributes of these elements. The general operations of a GNN can be broadly summarized in three main steps:

1) *Pre-processing:* an optional initial step that is typically performed offline for purposes such as sampling the graph, rearranging the graph to simplify the algorithm's processing and complexity, or encoding the feature vectors.

2) *Iterative updates:* the step where the main GNN computations occur through two main phases: aggregation and combination. The aggregation phase accumulates all the edges in a graph, and then for each vertex, it reduces all its neighbors and its own feature vectors into a single set. This feature set is combined and through performing linear transformations and non-linear activation functions, a new updated feature vector for each vertex is obtained. GNNs can be composed of several layers and the iterative process in a single layer updates every edge and vertex with information received from immediate neighbor vertices. This means that the relationships with nodes and edges that are progressively farther away can be gradually considered as more layers are processed.
3) *Readout*: the final step employed when a graph possesses a global feature vector, and it is updated once after the edge and node updates have been executed, usually in graph classification tasks.

Figure 1 illustrates an example of processing the first layer in a GNN. As shown, the aggregation phase iteratively gathers the neighbors of each vertex and then reduces all data into a single vector, $h_{vi}^a$. The reduce operation in this phase can be a variety of arithmetic functions, e.g., summation, mean, or maximum. This vector is then passed through the combination phase, which usually involves a neural network. Unlike conventional ANNs, where each layer has a different set of weights, vertices in a GNN all share the same weights.

Since GNNs were initially introduced in [11], multiple GNN algorithms and models have emerged. Graph Convolution Network (GCN) [12] expands the idea of convolution in the graph space. In contrast to CNNs, where the convolutional operation is defined on regular grid-like data, the convolutional operation in GCNs is defined on irregular graph structures. GraphSAGE [13] and Graph Isomorphism Network (GIN) [14] are two models that are also based on graph convolutions. GraphSAGE employs custom sampling techniques to obtain a fixed number of neighbors for each vertex while GIN learns the isomorphism invariant representation of graphs by using a learnable parameter $\varepsilon_l$ to adjust the weight of the central vertex. Graph Attention Networks (GATs) [15] are another class of GNNs that have demonstrated noteworthy results. GATs update node features through a pairwise function between nodes, which incorporates learnable weights. This results in an attention mechanism that can determine the usefulness of the edges.

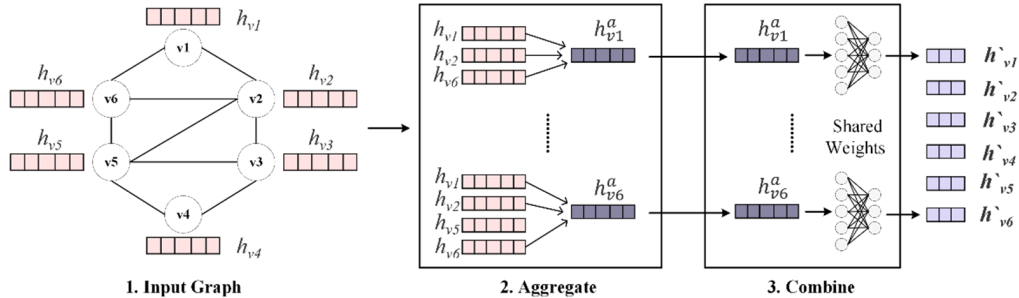

Figure 1: An example of GNN inference showing 1) Input graph to be processed; 2) Aggregation phase, where each vertex's neighbors are reduced to one feature vector; 3) Combine and Update phases, where each vertex is linearly transformed and updated using a non-linear activation function.

## 2.2 Graph neural network acceleration

Processing GNNs presents many challenges. A system processing a GNN needs to have the capabilities to efficiently handle both dense and very sparse computations, adapt its execution and operations based on the specific input graph structure and the GNN algorithm variant employed, and scale effectively to extremely large graphs. Due to the irregularity and large size of most real-world graphs, GNNs often require very high memory bandwidth and multiple irregular memory accesses. Further, the unique combination of computing characteristics from deep learning and graph processing in GNNs results in having alternate execution patterns [5]. Such challenges are typically absent when processing traditional ANN models. Thus, utilizing ANN accelerators for GNNs can be inefficient and lead to low performance and high energy costs. While overcoming these challenges is a non-trivial task, many recent efforts have tackled this problem and advanced the field of GNN processing, as discussed below.

On the software side, several frameworks and graph libraries have been proposed to aid in the acceleration of GNN models. A few examples of relevant libraries are PyTorch Geometric [16], Deep Graph Library [17], and NeuGraph [18]. Several programming models that aim to abstract GNN operations have also emerged, such as SAGA [18] and GRETA [19].

On the hardware side, multiple electronic hardware accelerators for GNNs have been proposed. The accelerator in [20] presents a modular architecture where the core unit of the accelerator is a tile composed of an Aggregator module (AGG), a DNN Accelerator module (DNA), a DNN Queue (DNQ), and a Graph Processing Element (GPE). The main component in their AGG module is a bank of ALUs, and the DNA exhibits the architecture of existing spatial accelerators. Another electronic accelerator EnG [21] has a unified architecture that handles GNNs in a single dataflow as a concatenated matrix multiplication of feature vectors, adjacency matrices, and weights. An array of clustered PEs is utilized. To aggregate the results, each column of PEs is connected in a ring, and the results are passed along and added based on the adjacency matrix. HyGCN [22] is another electronic accelerator for GCNs, composed of two dedicated engines that handle the aggregation and combination stages, along a control mechanism that coordinates the sequential execution of both processes. The dense combination stage is computed using a conventional systolic array approach. In contrast, the aggregation stage has a more complex architecture that includes a sampler, an edge scheduler, and a sparsity eliminator. Lastly, GRIP [23] utilizes the GReTA programming model [19] to create an electronic accelerator with specialized units and accumulators for edges and vertices, which are separate and adaptable.

Several GNN accelerators based on ReRAM and Processing-In-Memory (PIM) have also been presented. For example, ReGNN [24] leverages Analog PIM (APIM) and Digital PIM (DPIM). The authors decompose the computations in the combination phase into multiple Matrix-Vector Multiplications (MVM) and handle them through a dedicated combination engine composed of an APIM ReRAM array, while non-MVM operations are processed by the DPIM array. ReGraphX [25] is another ReRAM-based architecture that can be used for both training and inference acceleration of GNNs.

Unlike previous efforts, *GHOST* is the first GNN accelerator that leverages silicon photonics. It also supports accelerating a broad family of GNN models, adapts efficiently to different graph shapes and sizes, and mitigates typical GNN memory and performance bottlenecks.

**2.3 Silicon Photonics**

*2.3.1 Devices and Circuits*

Optical ANN accelerators have gained considerable attention from both academic researchers and industry in recent years because of their notable advantages in terms of energy-efficiency and performance [26]. There are two possible classes of optical ANN accelerators: coherent and non-coherent. In coherent architectures, a single wavelength is utilized to imprint parameters onto the optical signal's phase, which allows for Multiply and Accumulate (MAC) operations. Non-coherent architectures utilize multiple wavelengths and imprint parameters onto the optical signal's amplitude, enabling parallel operations to be performed using each wavelength. As opposed to conventional compute platforms such as GPUs and CPUs, silicon photonic CMOS fabrication does not require advanced technology nodes which mandate elevated process complexity, involving new lithography techniques and materials. Simple and less complex fabrication processes associated with older nodes are usually adopted instead [27]. The current focus of research in optical ANN accelerators is mainly on CNNs, MLPs, and RNNs. To the best of our knowledge, GHOST is the first optical accelerator for GNN models.

Figure 2 presents a general overview on the fundamental devices and circuits used for optical computing. The following are the main components needed:
- *Lasers*: used to generate optical signals that are needed to perform computation and communication in optical circuits. These lasers can either be on-chip or off-chip. While off-chip lasers have better light emission efficiency, there are significant losses when coupling the optical signals onto on-chip waveguides.

Conversely, on-chip lasers, such as vertical cavity surface emission lasers (VCSELs), offer a higher level of integration density and lower losses.

- *Waveguides*: silicon photonic waveguides carry the optical signal(s) generated by the laser source. They are typically composed of two materials, resulting in a high-refractive-index contrast such as a core made of Silicon (Si) and a cladding made of Silicon Dioxide (SiO2). This allows for total internal reflection. The waveguides can be either ridge or strip in shape. Wave Division Multiplexing (WDM) allows a single waveguide to support multiple wavelengths simultaneously without any interference. This enables the transmission of ultra-high bandwidth signals and is used for performing MAC operations.

- *Microring Resonators (MRs)*: An MR add-drop filter is an optical modulator which is designed using a ring-shaped waveguide. Each MR can be specifically designed and adjusted to work at a particular wavelength, known as the MR resonant wavelength ($\lambda_{MR}$), defined as $\lambda_{MR} = \frac{2\pi R}{m} n_{eff}$, where $R$ is the MR radius, $m$ is the order of the resonance, and $n_{eff}$ is the effective index of the device. Electronic data can be modulated onto the optical signal passing an MR by carefully adjusting $n_{eff}$ (and hence $\lambda_{MR}$) with a tuning circuit. MR banks consist of groups of MRs that share a single input waveguide and can be utilized for performing MAC and summation operations.

- *Photodetectors (PD)*: PDs are needed to detect processed optical signals and convert them into electrical signals. To be effective, a PD should be able to generate the desired electrical output using a small input optical signal. The input signal power from the laser source must be greater than the responsivity of the PD, while taking into consideration the different types of losses that may occur along the optical link.

- *Tuning circuits*: Tuning circuits are devices designed to control the effective index ($n_{eff}$) of MR devices to precisely modify an output optical signal. Typically, the tuning circuit is based on Thermo-Optic (TO) [28] or carrier injection Electro-Optic (EO) tuning [29], both of which cause a change in the effective refractive index ($n_{eff}$) and result in a resonant shift of $\Delta\lambda_{MR}$.

- *Digital-to-Analog Converters (DACs) and Analog-to-Digital Converters (ADCs)*: tuning the MR devices and converting the optical output to the electrical domain for intermediate buffering is all done using ADC and DAC devices. These devices represent one of the main performance bottlenecks in silicon-photonic-based systems due to their high latency and power costs. Accordingly, as will be discussed in Section 3, GHOST employs several techniques that aim to mitigate the downsides of these devices and reduce the needed opto-electric conversions.

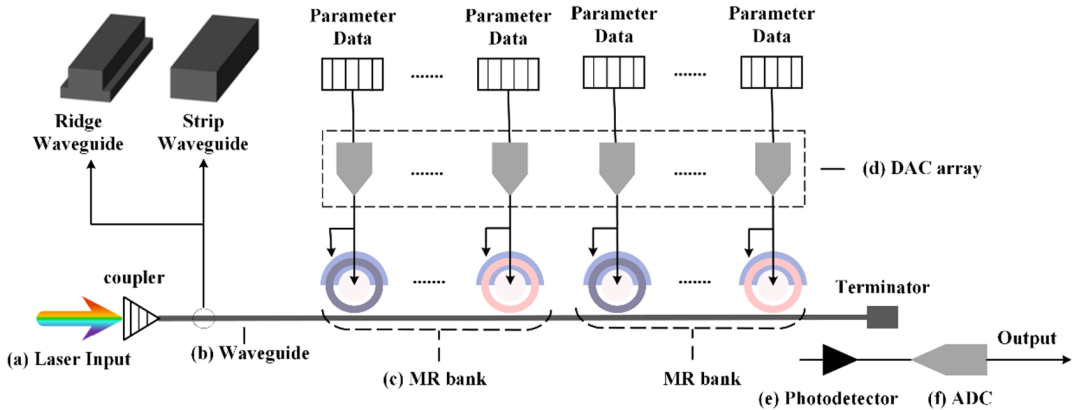

Fig. 2: Overview of photonic circuit used to implement operations for ANN acceleration. This circuit is composed of (a) a laser source, which can be off-chip or on-chip; (b) a waveguide, which can be strip (for passive devices) or ridge (for active devices); (c) MR banks perform MAC operation; (d) the banks are tuned as per data from the electronic domain, using DACs; (e) the result from the MAC operation is detected and accumulated using a PD; (f) for converting the data to digital domain, for postprocessing or storage, an ADC can be used..

### 2.3.2 Optical computation

Most of *GHOST's* core operations are performed using the opto-electronic tuning devices, MRs. Figure 3(a) shows a representation of the transmission plots for the input and the through ports' wavelengths after a parameter is imprinted onto the input signal. In most silicon-photonic-based systems, computations are performed as such by adjusting an MR's $\Delta\lambda_{MR}$, which leads to a predictable alteration in the amplitude of the optical signal's wavelength. GHOST leverages this to implement two main computations using MR devices: summation and multiplication. Summation is performed using coherent photonic summation. This entails using one optical signal with a single wavelength $\lambda_{MR}$ and MR devices adjusted to operate at the same resonant wavelength $\lambda_{MR}$. Figure 3(b) illustrates an example where coherent summation is used to add the values $a_1, a_2,$ and $a_3$. Using an analog biasing signal, VCSELs can be driven to produce an optical signal with a certain value imprinted onto it. Accordingly, the first value $a_1$ is imprinted onto the optical signal using the bottom VCSEL laser source. The top VCSEL produces an optical signal with a value of 1 which is split into two signals to be passed by the two MR devices. The first MR device then imprints the value $a_2$ onto the optical signal, while the second MR imprints the value $a_3$. When the optical signal generated by the bottom VCSEL unit ($a1$) and the one modulated by the first MR device ($a2$) meet, they undergo interference, resulting in a summation operation and an optical signal with the value $a1 + a2$ is generated. Similarly, when this optical signal meets the one modulated by the second MR device, they undergo interference, resulting in a summation operation and the final output $a1 + a2 + a3$ is computed. Coherent summation is ensured by using a laser phase locking mechanism [30], which guarantees that VCSEL output signals have the same phase for constructive interference to occur.

On the other hand, performing multiplications is done using non-coherent silicon photonics, where multiple optical signals with different wavelengths are multiplexed into the same waveguide using WDM. This enhances throughput and emulates neurons in ANNs as it involves combining multiple optical signals with different wavelengths into a single waveguide using an optical multiplexer [26]. Different wavelengths in the input waveguide pass through a series of MRs, with each MR tuned to a specific wavelength, allowing for several multiplications to be executed simultaneously in parallel. Figure 3 (c) illustrates an example of multiplying two vectors (activations vector $A$, and weights vector $W$) as follows

$$[a1 \quad a2 \quad a3] \times \begin{bmatrix} w1 \\ w2 \\ w3 \end{bmatrix} = a1w1 + a2w2 + a3w3 \qquad (1)$$

Two MR banks, and three optical signals with different wavelengths are needed to perform this multiplication. The first MR bank array imprints the activation values $a1, a2,$ and $a3$, while the second MR bank imprints the weight values $w1, w2,$ and $w3$. As shown in the figure, after imprinting the activation values, when the same signal with the same wavelength gets modulated by a second MR, a multiplication operation occurs between the previously imprinted value and the new one. Different optical wavelengths on the waveguide can then go through a PD to accumulate the output of the dot product. This method can be extended to perform MVMs since they can be decomposed into vector multiplications, by using several rows of the MR banks organization shown.

One of the main challenges of performing multiplications using non-coherent silicon photonics is the resulting heterodyne or incoherent crosstalk which is shown as the black portions in Figure 3(d). This mainly occurs when a portion of an optical signal from neighboring wavelengths interferes with the MR spectrum of another wavelength. In section 3.2, we discuss our efforts in optimizing the MR bank arrays design to reduce this crosstalk.

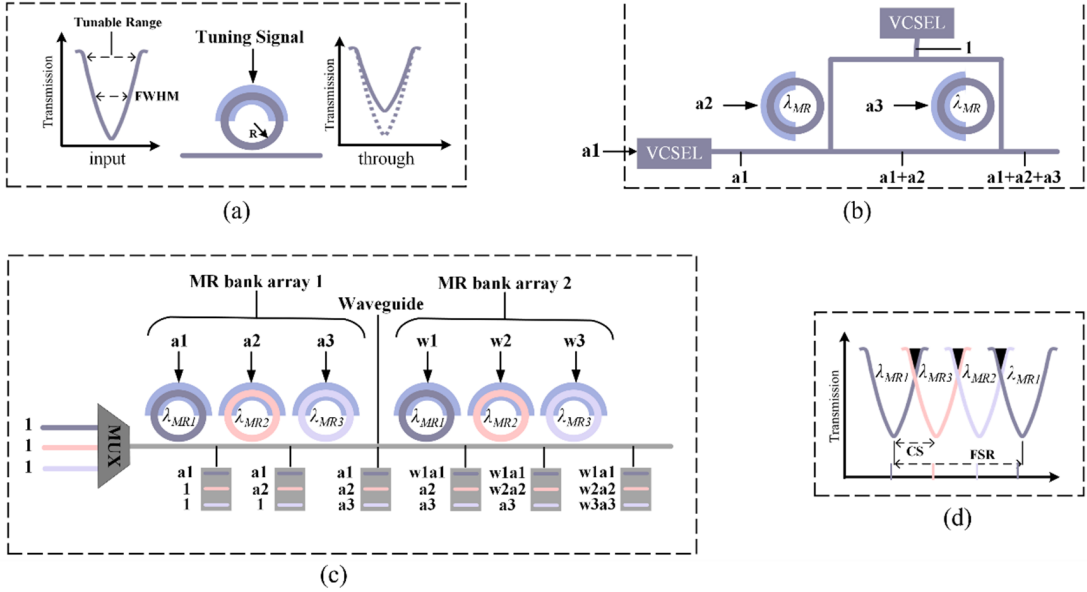

Figure 3: (a) MR input and through ports' wavelengths after imprinting a parameter onto the signal; (b) two MR devices used to perform optical coherent summation to add values $a_1$, $a_2$, and $a_3$; (c) MR bank arrays used to perform multiplication by imprinting input vector ($a_1$-$a_3$), followed by weight vector ($w_1$-$w_3$); (d) MR bank response and heterodyne crosstalk shown in black, where CS is channel spacing and FSR is free spectral range.

## 3 GHOST HARDWARE ACCELERATOR

*GHOST* is a silicon photonic architecture that can accelerate the inference of a diverse family of GNN models. An overview of the architecture is shown in Figure 4. The photonic accelerator core is composed of aggregate, combine, and update blocks, enabling the execution of a wide range of GNN models and real-world graph datasets. Interfacing with the main memory, buffering the input graph, identifying the needed resources, and mapping the weight matrices to the photonic architecture are all handled by an integrated Electronic-Control Unit (ECU). The following subsections describe the *GHOST* architecture and the device, circuit, and architecture layer optimization solutions we have considered to efficiently accelerate GNN models.

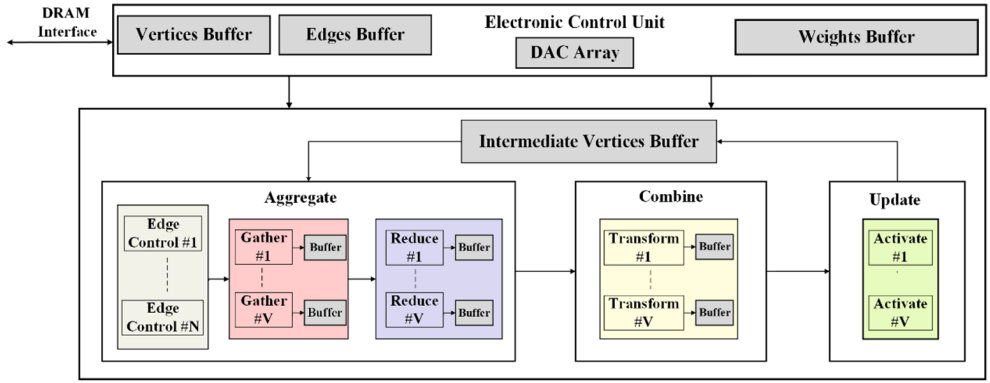

Figure 4: Overview of *GHOST* accelerator architecture showing the ECU, Aggregate, Combine, and Update blocks.

### 3.1 Tuning circuit design

As discussed earlier, MR devices require a tuning mechanism, which can be achieved using either EO or TO methods. In *GHOST*, we have implemented a hybrid tuning circuit that utilizes both methods to induce $\Delta\lambda_{MR}$. By doing so, we can capitalize on the advantages of each approach while mitigating their drawbacks. EO tuning is quicker (≈ns range) and consumes less power (≈4 μW/nm); however, it cannot be used for large tuning ranges [29]. Conversely, TO tuning offers a larger tunability range, but with the drawback of higher latency (≈μs range) and power consumption (≈27 mW/FSR) [28]. To address these challenges, we have integrated EO tuning to quickly induce small $\Delta\lambda_{MR}$ and reserved the slower TO tuning for cases where larger $\Delta\lambda_{MR}$ is required. The effectiveness of this hybrid approach has been previously demonstrated in [31]. To further reduce the power consumption of TO tuning and also reduce thermal crosstalk, we have employed the Thermal Eigenmode Decomposition method (TED) from [32]. We analyzed thermal interference between MR heaters and using Eigenmode decomposition, thermal tuning levels across heaters which do not cause thermal interference in MR banks were determined. Tuning the MR heaters according to the tuning levels obtained with this approach allowed us to mitigate thermal crosstalk and minimize TO tuning power.

### 3.2 MR device optimization

To ensure that the MVM operations performed are error free, so that the deployed GNN can be executed correctly, it is necessary to manage various sources of noise in the analog photonic domain. There are several major noise sources in the photonic computing substrate, including thermal crosstalk, heterodyne (or incoherent) crosstalk, and homodyne (or coherent) crosstalk. The thermal crosstalk between TO tuning circuits is mitigated using our TED-based tuning mechanism (Section 3.1). But we still have to mitigate the impact of heterodyne and homodyne crosstalk in our design.

The presence of multiple wavelengths (or channels) in the same waveguide causes heterodyne or inter-channel crosstalk, where a portion of an optical signal from neighboring wavelengths (say $\lambda_1$ and $\lambda_3$) can leak into the MR spectrum of another wavelength (say $\lambda_2$) (see Figure 3(c)). This power leak causes MR2 output to have $\lambda_1$ and $\lambda_3$ content, and the MR downstream (in this example MR3), will receive lower signal power than designed for. Thus, the output from MR2 and MR3 will be erroneous. For this design optimization, we model heterodyne crosstalk using the following equations:

$$P_{signal} = \Phi(\lambda_i, \lambda_j, Q_{factor}) P_s(\lambda_i, \lambda_j), \qquad (2)$$

$$P_{het\_noise} = \sum_{i=1}^{n} \Phi(\lambda_i, \lambda_j, Q_{factor}) P_s(\lambda_i, \lambda_j)(i \neq j), \qquad (3)$$

where $\Phi$ is the crosstalk coupling factor, i.e., the spectra overlap between the two neighboring wavelengths, $Q_{factor}$ is the quality factor or Q-factor of the MR, and $P_s$ is the input signal power to the MR.

Heterodyne crosstalk impacts signals with spectral overlap. To mitigate heterodyne crosstalk, this overlap should be minimized. This can be achieved by a well-designed channel spacing and Q-factor tuning while ensuring that the Signal-To-Noise ratio (SNR) in the output is higher than the photodetector sensitivity. Another factor to be considered is the tunable range of the designed MRs. The MRs should provide adequate Q-factor to improve SNR, but should also possess sufficient tunable range, i.e., 2×FWHM (FWHM=full width half maximum), so that necessary parameters can be imprinted error free. To mitigate heterodyne crosstalk, we optimize our design for high FWHM and SNR, where SNR is expressed as:

$$SNR = 10 \times \log_{10}(P_{signal}/P_{noise}), \qquad (4)$$

and, given $\lambda_{res}$ as the resonant wavelength of the MR being considered, FWHM can be modeled as:

$$FWHM = \lambda_{res}/Q_{factor}. \qquad (5)$$

Homodyne or coherent crosstalk is a result of undesired mode coupling among signals of the same wavelength [33]. In some of the computation circuits in *GHOST* we rely on coherent signal processing. In such circuits, part of the signal on the same wavelength may leak through a device and experience a different phase. Such leaked signals interfere with the output signal (based on their phase difference with the output signal) as coherent crosstalk noise. The presence of homodyne crosstalk, similar to heterodyne crosstalk, impacts the SNR of the non-coherent optical circuitry. The homodyne crosstalk noise power can be modeled as follows:

$$P_{hom\_noise} = \sum_{i=1}^{n} P_{in} \cdot X_{MR}^i(\rho) \cdot L_P^{n-i}, \qquad (6)$$

where $P_{in}$ is the input optical power, $X_{MR}^i(\rho)$ is the crosstalk contribution from the $i^{th}$ MR in a bank of $n$ MRs, and $\rho$ is the optical phase of the crosstalk signal, which is a function of the EO tuning voltage. $\rho$ does not take into account phase errors from thermal crosstalk, as TED is employed to address those errors. Finally, $L_P^{n-i}$ is the passing loss that the crosstalk signal experiences as it propagates through MRs in the coherent circuit.

For homodyne crosstalk mitigation we may increase the cross over coupling by increasing the gap between the input waveguide and ring waveguide. This reduces the amount of crosstalk signal being coupled over from the MR to the main waveguide, reducing the impact of crosstalk on the output signal. For achieving this, while meeting SNR constraints (discussed later) the $Q_{factor}$, attenuation in MR ($a$), and cross-over coupling coefficient ($\kappa$) has to be fine-tuned as follows [33]:

$$Q_{factor} = \frac{\pi n_g L \sqrt{(1-\kappa^2)a}}{\lambda_{MR}(1-a(1-\kappa^2))}. \tag{7}$$

Using our noise models described in (2), (3), and (6), we can identify the optimal design space for our MR banks which can ensure high SNR and a high tunable range ($R_{tune}$). We must also consider that the lowest optical power level ($P_{lpar}$) should be higher than $P_{noise}$.

$$P_{lpar} > P_{noise}, \tag{8}$$

$$\frac{P_{signal}}{P_{lpar}} < \frac{P_{signal}}{P_{noise}}, \tag{9}$$

$$10 \log_{10}\left(\frac{P_{signal}}{P_{lpar}}\right) < 10 \log_{10}\left(\frac{P_{signal}}{P_{noise}}\right), \tag{10}$$

where $P_{lpar}$ can be defined in terms of $P_{signal}$, from (2), as follows:

$$P_{lpar} = \frac{P_{signal} \times R_{tune}}{N_{levels}}. \tag{11}$$

Replacing $P_{lpar}$ in (10) yields the following relation:

$$10 \log_{10}\left(\frac{N_{levels}}{R_{tune}}\right) < SNR. \tag{12}$$

Here, $N_{levels}$ is the number of amplitude levels we need to represent across the available $R_{tune}$. For n-bit GNN parameter representation, $N_{levels}$ will be $2^n$. If positive and negative values are represented separately, as in the case with *GHOST*, then $N_{levels}$ will be $2^{n-1}$. The relationship in (12) can be rearranged to as follows:

$$\frac{2 \times \lambda_{MR}}{Q_{factor}} > N_{levels} \times 10^{-\frac{SNR}{10}} \tag{13}$$

Utilizing these models, we can identify the design space for our MRs and the MR banks they constitute, in terms of the $Q_{factor}$, $N_{levels}$, and $SNR$. We can obtain values for $\Phi(\lambda_i, \lambda_j, Q)$ in (1) and (2) and $X_{MR}(\rho) \cdot L_P^{n-i}$ in (5) through multi-physics simulations. The results from our exploration studies using our detailed models and the simulation tool suite from Ansys Lumerical [34] are presented later in Section 4.2.

### 3.3 GHOST architecture design

As illustrated in Figure 4, the main units in the *GHOST* architecture (aggregate, combine, update) are divided into $V$ execution lanes. During inference, each lane is assigned one output vertex to process, in parallel with all the other lanes. The aggregate block gathers all neighbor vertices and the associated edge data, and performs a reduce function for each of the assigned output vertices. The combine block then applies a linear transformation on each aggregated vertex feature vector $h_v^a$. Finally, the update block applies a non-linear activation function to obtain the updated vertex feature vectors $\grave{h}_v$. At the start of processing a GNN model, when processing the first layer, the aggregate block reads the graph data from the vertex and edge buffers in the ECU, which interface directly with main memory. As updated vertex data is computed, it is placed in the intermediate vertex buffer and thus, the aggregate block would read the vertex data when processing the next layers from the intermediate vertex buffer as needed.

### 3.3.1 Aggregate Block

As most graphs can be extremely sparse and irregular, regulating their memory accesses to improve performance is a challenge. *GHOST* alleviates this bottleneck through employing a "buffer and partition" optimization technique which is explained in Section 3.4.1. In this technique, the source and destination vertices in the input graph are split into blocks of $N$ and $V$. The aggregate block is thus composed of $N$ edge control units, $V$ gather units, and $V$ reduce units. In each cycle, the $V$ execution lanes in *GHOST* are assigned to one output vertex group at a time. The edge control units fetch $N$ input nodes simultaneously and then each unit forwards its fetched node and edge data to all gather units to convert the vertex data to analog signals, which are used to tune the MRs in the reduce units. The corresponding edge data is used by the gather units to define whether an input vertex is a neighbor of the assigned output vertex. Accordingly, the total delay of the aggregate block is dependent on the node with the largest number of neighbors.

The reduce unit is an optical unit configured as a coherent summation block. Each row in the reduce unit is assigned a feature from the vertex feature vectors, while each column is assigned a neighbor input vertex (see Figure 5(a)). The rows and columns have the sizes $R_r$ and $R_c$, respectively. Thus, one reduce unit can aggregate $R_r$ features of $R_c$ neighbor vertices. Different VCSEL and waveguide colors in the figure represent different optical wavelengths. Each VCSEL source generates an optical signal which is split into $R_c$ signals. These $R_c$ signals are then passed through the MR bank to imprint the neighbor nodes' feature values onto the signals. As explained in section 2.3, summation of the two values occurs when the different waveguides carrying signals with the same wavelength meet and they undergo interference. Since for a particular vertex the number of its neighbors may be greater than $Rc$, multiple mappings of different source vertices may be needed. Accordingly, the output of each row in the reduce unit is converted to an analog signal using a PD, and used to tune the last MR in each lane (Figure 5(a)) such that the sum that is output from that cycle will be added to the feature values in the next cycle.

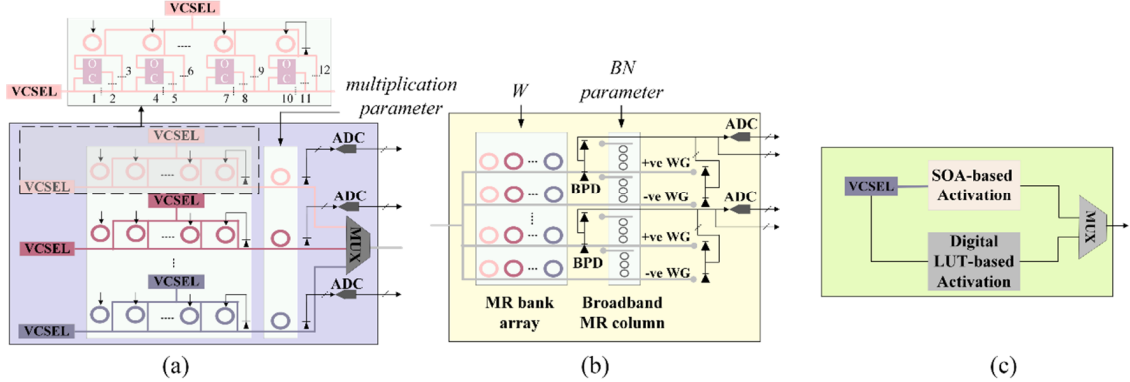

Figure 5: (a) Reduce unit showing the needed changes in each feature lane to support the max aggregation operation using an optical comparator; (b) detailed view of transform unit; (c) detailed view of activate unit.

The organization of the reduce unit allows for the flexibility of implementing a wide range of reduce operations, which encompass most, if not all, GNN models. After the summation step as explained above, the output of the reduce unit is $(h_v^a = h_v + \sum_{u=0}^{n} h_u)$. The last MR after the coherent summation block is used to implement the mean operation where the output summation value would be adjusted by the MR such that it is multiplied by $\frac{1}{number\ of\ neighbors\ (n)}$, resulting in a reduce unit output of $(h_v^a = h_v + \frac{1}{n}\sum_{u=0}^{n} h_u)$. Further, for implementing other reduce operations, such as maximum, the reduce unit includes an optical comparator [35]. An example of one lane in the reduce unit in that case would be as shown at the top of Figure 5(a). By activating blockers $1, 3, 4, 6, 7, 9, 10$, and $12$, the output of the reduce unit becomes the maximum value among all nodes. The optical output from all the rows in the reduce unit would then be combined into a single waveguide. This optical waveguide is connected directly to transform units in the combine block (discussed in the next subsection) to undergo the needed linear transformations. The same output values may need to be passed to the transform unit multiple times, depending on the size of the weight matrix. Accordingly, as the output from the reduce unit is passed directly to the transform units, it will be converted to the digital domain and buffered.

### 3.3.2 Combine Block

The combine block accumulates the results from the aggregate block and performs linear transformation using the learned weight parameters. This generates a new learned, more expressive representation that captures the important structural information of the graph. Additionally, the linear transformation usually results in a reduced dimensionality for the vertex data representation, making the model more efficient as the dimensionality of some graphs can be very large (as shown later in Table 2). The transform unit's linear transformation is computed in the optical domain using MR bank arrays, as shown in Figure 5(b). Since linear transformation operations in GNNs are mainly MVMs, where the feature vector of the vertex being computed is multiplied by the weight matrix, as discussed in Section 2.3.2, such operations can be computed in the optical domain using MR bank arrays by passing the weight values to the transform unit as analog signals to tune each MR, while the feature vectors values are imprinted onto the optical signals in the waveguide from the reduce units. Unlike the MR array used in reduce units to perform summation operations, as presented in Subsection 3.3.1, multiplications are computed using non-coherent silicon photonics where multiple optical signals with different wavelengths are multiplexed using WDM into a single waveguide and each MR is adjusted to operate at one of those wavelengths used. Accordingly, each feature value output from the reduce units and imprinted onto one optical signal's wavelength is multiplied by the weight values in the transform units. The number of rows in a reduce unit $R_r$ is equal to the number of columns in a transform unit such that the number of optical signals in the waveguide is equal to the number of MRs in one row in the transform unit. The number of columns in a transform unit however is $T_r$.

As Batch Normalization (BN) is commonly used in many GNNs after the linear transformation, *GHOST* leverages broadband MR devices to perform BN in the optical domain where the BN parameter is used to tune the broadband MRs to adjust the optical signals as needed to reflect the BN operation. The efficiency of performing BN optically with this configuration was demonstrated in [35]. It is important to note that the BN unit can be bypassed if not needed. The output from each row is then accumulated using Balanced Photodetectors (BPD). BPDs are photodetectors that have two separate arms for the same waveguide, one for positive and the other for negative signal polarities. This allows them to accommodate both positive and negative parameter values by detecting the absolute difference between the two signals. The BPD sums the output signal from the positive arm and the output signal from the negative arm separately. Then, the BPD subtracts the output signal from the negative arm from the output signal from the positive arm to obtain the net difference signal.

If the size of rows in the weight matrix is smaller than or equal to the number of columns in the transform bank array and only one mapping for each weight matrix row is needed, the output from the transform unit after the BPDs will be passed directly to the activate units. Otherwise, the output will need to be converted to the digital domain and buffered till all needed values are computed and accumulated, and then passed to the activate units in the update block (discussed in the next subsection). *GHOST*'s versatility to adapt to different model architectures and sizes, and not having to always convert the values to the digital domain, greatly reduces the latency and power costs associated with Analog-to-Digital Converters (ADCs) and buffering.

### 3.3.3 Update Block

The update block is composed of $V$ update units to apply a non-linear activation function to the output from the transform units. The work in [36] demonstrated how semiconductor-optical-amplifiers (SOAs) can be exploited to implement multiple non-linear functions such as *RELU*, *sigmoid*, and *tanh*. For example, when the gain in an SOA is adjusted to a value close to 1, the behavior resembles the *RELU* operation. Accordingly, such non-linear operations are implemented optically, resulting in considerably improved performance. The analog signals output from the transform units are used to directly drive VCSELs, generating optical signals with the output value imprinted into its amplitude, which is then passed through the SOA-based non-linear unit. For the non-linear activation functions that are harder to implement optically (such as softmax), a digital activation unit, such as the one described in [37] can be integrated to accommodate these functions using Look-Up Tables (LUTs) and simple digital circuits, such as add and subtract. One update unit consists of $T_r$ rows of activate units, in compliance with the number of rows used in each transform unit. Accordingly, the output from each row of the transform unit corresponds to one value in the vertex data vector, as shown in Figure 5(c).

### 3.4 Orchestration and scheduling optimizations

*GHOST* supports four main optimizations for efficient orchestration and scheduling of GNNs, namely 1) graph buffering and partitioning, 2) execution pipelining and scheduling, 3) weight DACs sharing, and 4) workload balancing. While performing computations in the optical domain already offers significant performance and energy benefits, efficient optimizations targeting improved memory bandwidth utilization and enhanced execution flow are imperative for designing a scalable and robust GNN accelerator.

#### 3.4.1 Graph buffering and partitioning

Retrieving the entire graph from memory and processing it all at the same time entails tremendous memory bandwidth, resources, and computational costs. Hence, dividing the graph into several partitions and executing them separately is a widely used GNN optimization. But utilizing this approach alone can lead to increased latency as, in the worst case, while processing one vertex, it might necessitate loading another partition for each of its neighbor vertices. *GHOST*, on the other hand, uses a modified partitioning algorithm that builds on the one presented in [23]. As information regarding the graph edges is input, the adjacency matrix for the graph is generated. Columns of the adjacency matrix are identified as destination/output vertices while rows are the source/input vertices. Output and input vertices are then partitioned into groups of sizes $V$ and $N$, respectively. Consequently, the edges data are also grouped into $V \times N$ chunks. For each partition where the group $V_i$ is assigned to the vertex processing lanes, if for input nodes $N_i$, the corresponding group of edges contains one or more connected edges, $N_i$ is prefetched and assigned to the edge control units, while all-zero blocks are skipped entirely. Generating the partition matrices and determining the memory access and fetching order is done once offline, as part of a graph preprocessing step. This significantly reduces sparsity in the graph and the need for complex techniques to deal with performing sparse operations. It also helps overcome the GNN's memory bottleneck and eases greatly the traffic and access frequency to and from the main memory.

#### 3.4.2 Execution pipelining and scheduling

*GHOST* can efficiently adapt the pipelining, computation scheduling, and execution ordering depending on the GNN model, as each model can require a different sequence of execution. For example, GCNs usually mandate having all nodes gathered first, reduced, transformed, and then updated. In contrast, GATs initially compute the attention coefficients, which involves gathering the nodes, performing linear transformations, and then applying a non-linear activation. In this case, the reduce step is performed at the end.

Given the buffering and partitioning optimization discussed in Section 3.4.1, *GHOST* performs pipelining at two levels of granularity. The first pipelining technique entails pipelining the operations within one output vertex group $V_i$. Initially as the output vertices are assigned to the execution lanes, all of their neighbor nodes are gathered. In case of models such as GCN, GraphSAGE, and GIN, which perform aggregation first and then the transform and update, *GHOST* pipelines the reduce, transform, and update executions. As soon as $R_c$ (number of columns in a reduce unit) input vertices are gathered, the reduce units are initiated. Transform units are activated when $R_r$ (number of rows in a reduce unit) feature values are ready and output from the reduce units, without the need to wait for all feature values to be computed. Similarly, the update units begin updating the vertex data directly after $T_r$ (number of rows in a transform unit) values are linearly transformed. For the second pipelining technique between different vertex groups, the operations for a group $V_{i+1}$, are pipelined with those of group $V_i$. This scheduling approach ensures that the initial reduce for $V_{i+1}$ is activated after the last reduce for group $V_i$. The pipelining model is illustrated in Figure 6(a).

For other GNN models such as GATs, where transforming the vertex data occurs prior to the aggregation, the pipelining model is tailored to fit the model's specifications. Figure 6(b) shows an example of pipelining for a GAT model. For pipelining with one output vertex group, as the transformation of each vertex fetched by the edge control units is independent of the other vertices, gather operations are pipelined with the transform operations without the need to wait till all needed partitions are fetched and processed. Hence, the first group of input vertices is gathered, and the first linear transformation of multiplying the input with matrix $W$ is handled by transform units, followed by attention vector multiplication. The output is then concatenated with transformed output vertices updated with a *leakyRELU* activation function by the activate units. When all the neighbor vertices $\in V$ are transformed, softmax is computed. Reduce is then computed at the end after the second transformation. Alternatively, when processing all output vertex groups $V$, the gather operations for the next output vertex group $V_{i+1}$ are pipelined with the transform and update operations for vertex output group $V_i$.

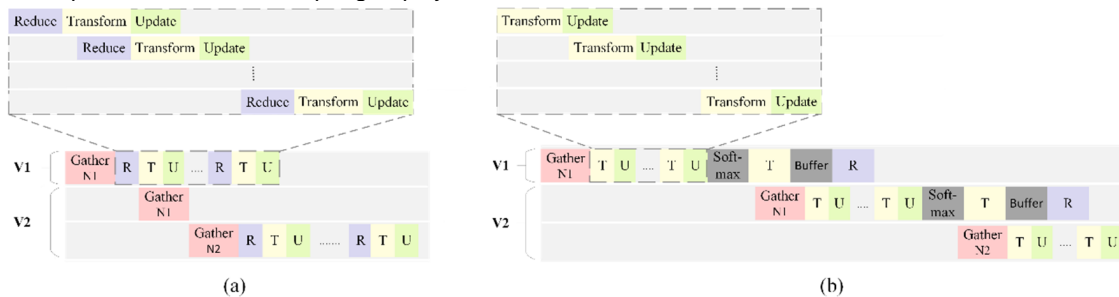

Figure 6: *GHOST* pipelining within one output vertex group $V_1$ (top), and the pipelining between output vertex groups $V1$, $V2$, where $V1$ requires input vertices in $N1$ while $V2$ requires $N1$ and $N2$ for (a) GCN, GraphSAGE, GIN; and (b) GAT.

### 3.4.3 Weight DAC sharing

A key characteristic of GNNs is that the same set of weights is applied across all vertices during processing. Accordingly, when all transform units in *GHOST* are operating with the same speed and processing vertices with the same feature sizes, the same weight values can be shared among all transform units. *GHOST* leverages this idea to implement a weight DAC sharing optimization. DAC devices are needed in *GHOST* to convert the digital values of weights that are read from the buffers in the ECU into analog signals. These analog signals are then used to tune optical MR devices to perform the linear transformations in each transfer unit. The DAC sharing optimization shares DACs across weights, thereby reducing the total number of DAC devices required, which is a significant factor in the power and latency budget of silicon-photonic accelerators, as normally one DAC device would be needed for each MR. With DAC sharing $V$ (number of transform units) MRs would share the same DAC device and thus the total number of DAC devices is reduce by $\frac{\#MRs\ in\ combine\ block}{\#MRs\ in\ one\ transfer\ unit}$.

### 3.4.4 Workload balancing

Exploiting the buffer and partitioning optimization discussed in Section 3.4.1 implies having certain units/blocks in idle states during specific times. Due to some graphs' irregularity, the number of neighbors for each vertex within the same output vertex group $V_i$ can vary considerably. As a result, when processing a GCN model for example, some gather units will be waiting in an idle state till the gather unit with the highest dimensionality vertex receives all its neighbor vertices. Our workload balancing optimization allows each lane to operate at its own rate without the need to wait while other lanes with higher dimensionality are still gathering their needed vertices. Accordingly, when such lanes complete processing their assigned vertices, the workload of the other lanes will be split among the completed lanes. As a result, this can notably reduce the overall latency of the executing GNN model, especially when dealing with highly sparse and irregular graphs.

### 3.5 Programming model

*GHOST*'s programming model is based on GReTA [19], an abstraction model tailored to processing a broad family of GNNs. GReTA utilizes four stateless User-Defined Functions (UDFs) to break down computation in each GNN layer, namely gather, reduce, transform, and activate. These UDFs are then performed in a series of three main execution phases: aggregate, combine, and update. Algorithm 1 explains the execution flow of the GReTA programming model as implemented by *GHOST*. *GHOST* takes as input the graph(s) as a set of edges and vertices represented by the notation *G(V,E)*, where edges are defined as two lists of vertices: destination/output ($V$), and source/input vertices ($U$).

The aggregate phase iterates over all edges and invokes gather and reduce UDFs. The *Gather()* UDF collects each output vertex feature vector ($h_v$), its neighbor input vertices ($\forall h_u \in N(v)$) and the edges features ($h_{u,v}$) associated with the current output vertex being processed, and prepares a message value. The *Reduce()* UDF collects the messages that are output by gather, related to the same output vertex, and reduces them into a single value. The combine phase invokes the *Transform()* UDF where it takes in all the accumulated values for each vertex, employs the learned weight parameters, and performs linear transformation. Lastly, the update phase invokes the *Activate()* UDF and computes a non-linear activation function for each output vertex to generate the updated feature vectors for all vertices.

ALGORITHM 1: GHOST Programming Model

**Input:** Graph *G(V, E)*, source vertex feature data $h_v$, vertex feature data $h_u$, edges feature data $h_{u,v}$, weights *W*.
**Output:** Updated vertex feature data $h_v'$
1: // Edges Accumulate Phase
2: **for** each (u, v) in E:
3:     $h_{v\_r}$ = Reduce($h_v$, Gather($h_u$, $h_v$, $h_{u,v}$))
4: // Vertices Accumulate Phase
5: **for** each v in V:
6:     $h_{v\_t}$ = Transform($h_v$, W)
7: // Update Vertices Phase
8: **for** each v in V:
9:     $h_v'$ = Activate($h_{v\_t}$)

## 4 EXPERIMENTAL RESULTS

### 4.1 Simulation setup

To evaluate our proposed *GHOST* accelerator, we developed a comprehensive simulator in Python to estimate the power and latency of the accelerator. *GHOST* was simulated with attention to both software mapping and hardware mapping. For the software mapping, graph data is used to generate the partition matrix and retrieve needed information about the graph, such as the maximum number of neighbors in each partition. Then, we consider the layer-wise mapping and operation of each GNN model, and the architectural requirement for the mapping. For the hardware mapping, we modeled optoelectronic and electronic devices and circuits, and composed these into the blocks of the accelerator architecture. Compact models were used to analyze the losses associated with the device operation and also to determine device latency. The performance and energy estimates of all buffers used in *GHOST* were obtained using CACTI [38]. However, since CACTI only supports down to 20 nm technology, the obtained latency and energy values were scaled down to 7 nm using the set of scaling relations from [40]. The off-chip DRAM memory considered employs HBM2 [41] with a size of 8GB, and it was simulated using DRAMsim3 [39]. The maximum bandwidth required to accommodate the largest graph dataset across all the different GNN models used in our experiments is 174.4 GB/s (the HBM2 memory system can support a maximum bandwidth of 256 GB/s [41]). The ECU is composed of 4 main buffers: input vertices buffer (128KB), output vertices buffer (128KB), edges buffer (256KB), and weights buffer (128KB). The memory bandwidth and access latencies of off-chip memory and on-chip buffers have all been accurately modeled using the methodology mentioned and are considered in all our simulations. For the softmax circuit needed for the GAT model using the update block, the design with a LUT and a maximum frequency of 294 MHz from [37] was utilized.

Table 1 displays the optoelectronic device and circuit parameters that were used during *GHOST*'s simulation-based analysis. Various factors were taken into account for assessing photonic signal losses, including waveguide propagation loss (1 dB/cm), splitter loss (0.13 dB [42]), combiner loss (0.9 dB [42]), MR through loss (0.02 dB [44]),

MR modulation loss (0.72 dB [45]), EO tuning loss (6 dB/cm [29]), and TO tuning power (27.5 mW/FSR [28]). Also, as the number of wavelengths and waveguide length increase, so does the MR count, photonic loss, and required laser power consumption. Thus, the laser power required for each source used with multiple wavelengths in our architecture is modeled as follows:

$$P_{laser} - S_{detector} \geq P_{photo\_loss} + 10 \times \log_{10} N_\lambda, \tag{13}$$

where $P_{laser}$ is the laser power (dBm), $S_{detector}$ is the PD sensitivity (dBm), $N_\lambda$ is the number of laser sources/wavelengths, and $P_{photo\_loss}$ is the total optical loss (dB) due to the factors discussed.

Table 1: Parameters considered for GHOST analysis

| Devices | Latency | Power |
|---|---|---|
| EO Tuning [29] | 20 ns | 4 µW/nm |
| TO Tuning [28] | 4 µs | 27.5 mW/FSR |
| VCSEL [10] | 0.07 ns | 1.3 mW |
| Photodetector [10] | 5.8 ps | 2.8 mW |
| SOA [10] | 0.3 ns | 2.2 mW |
| DAC (8 bit) [46] | 0.29 ns | 3 mW |
| ADC (8 bit) [47] | 0.82 ns | 3.1 mW |

Table 2: Graph Datasets Parameters

| Dataset | (avg) #Nodes | (avg) #Edges | #Features | #Labels | #Graphs |
|---|---|---|---|---|---|
| Cora | 2,708 | 10,556 | 1,433 | 7 | 1 |
| PubMed | 19,717 | 88,651 | 500 | 3 | 1 |
| Citeseer | 3,327 | 9,104 | 3,703 | 6 | 1 |
| Amazon | 7,650 | 238,162 | 745 | 8 | 1 |
| Proteins | 39 | 73 | 3 | 2 | 1113 |
| Mutag | 18 | 40 | 143 | 2 | 188 |
| BZR | 34 | 38 | 189 | 2 | 405 |
| IMDB-binary | 20 | 193 | 136 | 2 | 1000 |

Table 3: GNN Models Performances

| Model | Dataset | Accuracy (32-bit) | Accuracy (8-bit) |
|---|---|---|---|
| GCN | Cora | 88.70% | 88.90% |
| | PubMed | 87.40% | 87.30% |
| | Citeseer | 74.90% | 74.40% |
| | Amazon | 94.00% | 93.70% |
| GraphSAGE | Cora | 71.70% | 70.80% |
| | PubMed | 77.50% | 76.90% |
| | Citeseer | 63.30% | 65% |
| | Amazon | 77.10% | 76.90% |
| GAT | Cora | 78.30% | 77.90% |
| | PubMed | 76.70% | 77.90% |
| | Citeseer | 70.20% | 69.10% |
| | Amazon | 94.38% | 94.64% |
| GIN | Proteins | 74.00% | 73.40% |
| | Mutag | 94.74% | 94.74% |
| | BZR | 65.85% | 65.85% |
| | IMDB-binary | 77% | 73% |

A diverse set of GNN models, tasks, and graph datasets were used in our analysis. The following GNN models were considered: GCN [12], GraphSAGE [13], GIN [14], and GAT [15]. Each model processed four different graph datasets with the properties outlined in Table 2. The node-classification graph datasets (Cora, PubMed, Citeseer, Amazon) were processed with GCN, GraphSAGE, and GAT, while GIN was used for processing the graph-classification datasets (Proteins, Mutag, BZR, IMDB-binary). Further, GCN and GraphSAGE were implemented with two layers, while the MLP in GIN was implemented with eight layers. For the GAT model, two layers were implemented with the first one leveraging eight attention heads while the second used one attention head. The PyTorch Geometric library [16] was used to train and analyze each model's accuracy as shown in Table 3. Our analysis indicated that 8-bit model quantization results in comparable algorithmic accuracy to models with full (32-bit) precision; thus, we targeted the acceleration of 8-bit precision GNN models.

In the following subsections, we present results from our analyses and experiments to determine optimal photonic device level configurations in GHOST (Subsection 4.2), the optimal values for GHOST's architectural parameters, which were discussed in Section 3.3, including $V$, $N$, $R_r$, $R_c$, and $T_r$, (Subection 4.3), sensitivity analysis to assess the impact of the orchestration and scheduling optimizations that were discussed in Section 3.4 (Subsection 4.4), and comparison with GNN accelerators proposed in prior work (Subsection 4.5).

### 4.2 Device-level analysis

To ensure error-free photonic device operation in GHOST, we must reduce various noise sources in the analog domain, and ensure higher SNR than that dictated by (11) (see Section 3.2). We performed our device-level optimization analysis using optoelectronic simulation tools from Ansys Lumerical [34]. For obtaining the operational characteristics of the active MRs, i.e., MRs with EO tuning as opposed to passive MRs without them, we used the FDTD, CHARGE, MODE solver, and INTERCONNECT tools [34]. FDTD was used to obtain the operational characteristics of the passive MR. The doping levels for the tuning junction's p and n doped regions was assumed to be 4e19 in our CHARGE simulations, to obtain the carrier distribution in the ring, under various voltage conditions (0V to 10V range). Using the carrier distribution to voltage relationship obtained from CHARGE, we performed MODE solver simulations to obtain the shift in effective refractive index ($n_{eff}$) over the voltage range. Finally, the data from these simulations was used in our INTERCONNECT simulations to obtain the operational characteristics of the MR under different biasing voltages. These analyses were performed over a range of design parameters for the MR and $\lambda_{MR}$ values. From these simulations, we obtained our ideal MR design to have: ring and input waveguide width at 450 nm, radius of 10 μm, Gap of 300 nm, and a Q-factor of 3100. Using these parameters and (12) we can calculate the SNR required to be 21.3 dB.

The data from these tools was used for crosstalk and SNR analysis as mentioned previously in Section 3.2. Depending on the SNR requirements, the gap between the input and the ring waveguide and the width of the waveguides (input and ring) were explored and adjusted to achieve the required trade-off between Q-factor and the SNR at the output of the design as per (12). N_levels in our design is fixed from our software-level quantization to be $2^7$, since we consider 8-bit quantization in our models (see Section 3.2 for discussion on this).

Using the device operational characteristics and the noise models from Section 3.2, we perform a sweep to determine the viable MR bank sizes for our coherent circuits ($P_{noise} = P_{\text{hom}\_noise}$) and non-coherent circuits ($P_{noise} = P_{het\_noise} + P_{\text{hom}\_noise}$), the results of which are shown in Figures 7(a) and 7(b).

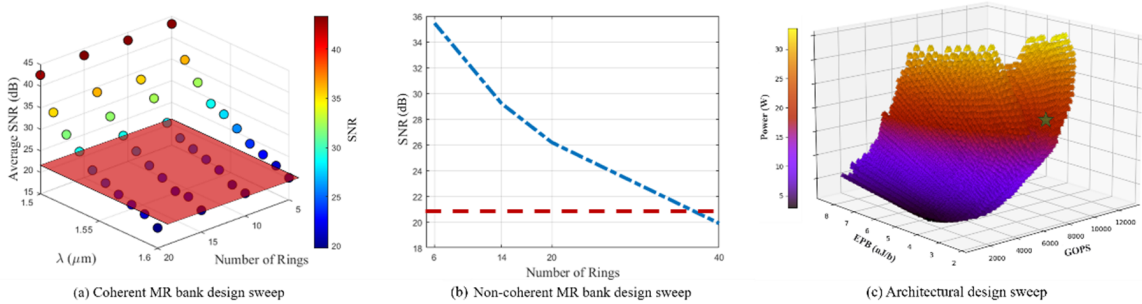

Figure 7: Design space exploration for (a) coherent and (b) non-coherent MR banks for *GHOST* architecture; (c) Architectural design-space exploration for *GHOST*, to find the optimal $[N, V, Rr, Rc, Tr]$ configuration with the best EPB/GOPS. The best configuration, [20,20,18,7,17] is shown with the green star.

For coherent MR banks we need to sweep for the wavelength to be used in the circuit, the number of MRs, and the SNR for the design. The cutoff SNR is shown as a red plane in the figure. From Figure 6(a), we can observe that it is possible to have up to 20 MRs in the coherent summation circuit, when the resonant wavelength is 1520 nm, while satisfying the SNR requirements (red plane). Similarly, exploration for non-coherent circuits was also conducted and results are shown in Figure 7(b). The number of rings (MRs) along the x-axis is 2 times the number of wavelengths in the waveguide, as we need two MR banks to perform multiply and accumulate operations. We considered the first wavelength to be 1550 nm and used a channel spacing of 1 nm between wavelengths. The red line in Figure 7(b) indicates the cut-off SNR. From this analysis, we determined that the waveguide can host 36 MRs, or 18 wavelengths (1550 nm to 1568 nm) for non-coherent operation. These results were used to size and design the MR banks within the main architectural blocks of *GHOST*, to meet SNR goals while maximizing performance.

### 4.3 Architecture design space exploration

The *GHOST* architecture relies on five main parameters, as outlined in Section 3: $N, V, R_r, R_c$ and $T_r$. $N$ refers to the number of edge control units or the size of each input vertices group in the partition matrix, while $V$ refers to the number of execution lanes, which is also the size of each output node group in the partition matrix. $R_c$ is the number of columns and $R_r$ is the number of rows in the coherent sum MR array in the reduce units, which is also the number of columns for the MR bank array in the transform units. Lastly, $T_r$ is the number of rows for the MR bank array in the transform units. To identify the optimal configuration for *GHOST*, which is determined by the combination of $[N, V, Rr, Rc, Tr]$ that offers the lowest EPB/GOPS (where EPB is energy-per-bit and GOPS is giga-operations-per-second), we conducted a detailed design space exploration as shown in Figure 7(c). Using the *GHOST* simulator described in Section 4.1, the EPB and GOPs values were obtained for each GNN model and each accompanying graph dataset for a wide set of possible values for [$N, V, Rr, Rc, Tr$]. The average EPB/GOPs values across all the GNN models and datasets for each set of parameters were then obtained and the optimal configuration [20,20,18,7,17] was identified as the one with the lowest EPB/GOPS value.

### 4.4 Orchestration and scheduling optimization analysis

We conducted a sensitivity analysis to assess the impact of each of the orchestration and scheduling optimizations described in Section 3.4. The normalized energy results are shown in Figure 8. The baseline configuration does not utilize any of the optimizations and each gather unit requests the needed neighbor vertices sequentially from the ECU. The Buffer and Partition (BP), Pipelining (PP), and DAC weight sharing (DAC_Sharing) optimizations and their viable combinations were explored in this analysis. For Workload Balancing (WB), implementing it in isolation was found to inefficient as the memory and buffer accesses are not synchronized and occur sequentially and on-demand. For instance, the processing lane that handles the vertex with the smallest dimensionality may not be the first to finish execution, as the order of memory access is also a critical factor. Moreover, employing WB necessitates having each lane possibly operating at different speeds, making it difficult to utilize the weight DAC sharing optimization. Therefore, implementing WB in isolation is impractical, and we only show results of considering WB alongside BP and PP to observe its benefits.

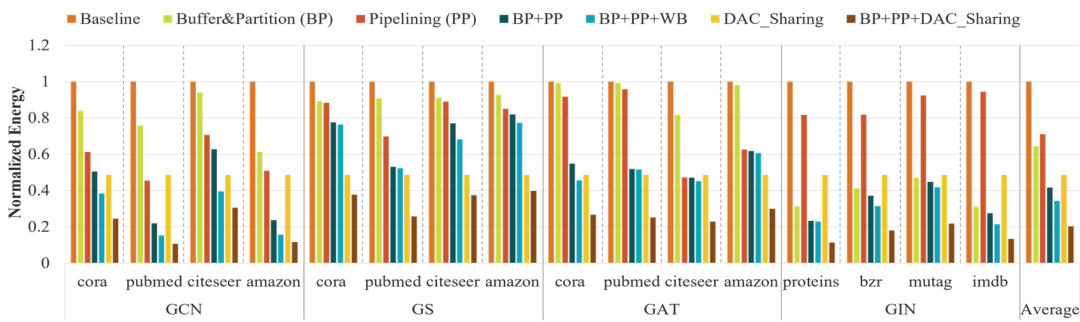

Figure 8: Impact of each orchestration and scheduling optimization on normalized energy consumption in *GHOST*.

The results shown in Figure 8 are normalized to the energy consumption of the baseline model. As can be observed, employing BP, PP, and DAC sharing optimizations simultaneously results in the least energy values for all the GNN models as well as all the graph datasets used. On average, when using DAC sharing combined with BP and PP, the energy consumption is reduced by 4.94× compared to the baseline. On the other hand, using BP, PP, and WB reduces the overall energy consumption by 2.92×. Hence, while *GHOST* supports all four optimizations described in Section 3.4, we leverage BP, PP, and DAC sharing for our *GHOST* configuration that is used in the subsequent sections for comparisons with other GNN accelerators.

Figure 8 also indicates that the different optimization techniques have different impacts across the various GNN models and graph datasets used. The optimizations have a greater impact with datasets having large number of vertices, and a very high degree of sparsity (e.g., PubMed). This demonstrates the scalability of *GHOST* to larger and more complex graphs and how the optimization techniques can alleviate the expensive memory and computational costs associated with GNN inference.

Another key observation is the effect of BP and PP when processing different graphs. When processing larger graphs such as Cora, PubMed, Citeseer, and Amazon, PP results in lower energy consumption values than BP. Conversely, for datasets used with GIN, BP leads to lower energy values. While the graph datasets used with GIN are each composed of multiple graphs, each individual graph in the datasets is considerably smaller than the other graphs used with GCN, GraphSAGE and GAT. Accordingly, as pipelining is determined based on each individual graph, the impact of PP with small graphs diminishes. On the other hand, BP reduces sparsity and number of memory accesses. Consequently, as we need to offload an entire graph from memory every time *GHOST* starts processing a new graph, employing BP results in notable energy reductions.

### 4.5 GHOST architecture component-wise performance analysis

To understand the performance of the major blocks in the GHOST architecture, we present a breakdown in Figure 9 in terms of latency when processing each GNN model and graph dataset for each of the main blocks: aggregate, combine, and update. It is evident from Figure 9 that the performance and contribution of each block to the overall inference latency depends on the GNN model being processed and the graph dataset used. In general, aggregate consumes more than half of the latency budget when processing GCN and GS models. This is mainly due to the models operating on graph datasets with large feature vectors and relatively high node degrees. Accordingly, the aggregate phase takes longer time as all neighbor node groups from the partitions matrix (from the graph buffering and partitioning technique in Section 3.4.1) need to be fetched and added. On the other hand, while GAT processes the same graph datasets, it follows a different execution ordering and pipelining as explained in Section 3.4.2. The GAT model used in our experiments is composed of eight attention heads, which are computed using the combine block. Moreover, the softmax function performed by the update block and included in the GAT computations is more time-consuming than the non-linear activation RELU used in GCN and GS models. The aggregation is performed only once at the end. Consequently, the high latencies observed with processing GATs are mainly attributed to the combine and update phases. Lastly, the graph datasets processed by the GIN model are multiple graphs used for graph classification tasks. However, each single graph is much smaller and possesses lower node degrees than the other graphs used with GCN, GS, and GAT. Accordingly, since a smaller number of neighbor nodes need to be reduced, the aggregate phase is not as time consuming as with other models and the performance bottleneck can be attributed to the combine phase.

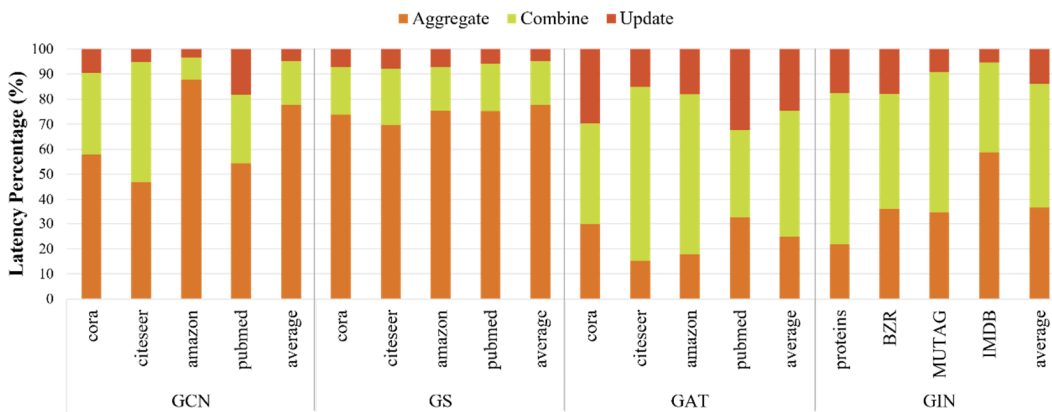

Figure 9: Breakdown analysis results showing the impact of each block in GHOST on the performance for each GNN model and graph dataset.

### 4.6 Comparison with computing platforms and state-of-the-art GNN accelerators

*GHOST* is compared against multiple computing platforms and state-of-the-art GNN hardware accelerators: GRIP [23], HyGCN [22], EnG [21], HW_ACC [20], ReGNN [24], ReGraphx[25], Google TPU v4, Intel Xeon CPU, and NVIDIA A100 GPU. We used power, latency, and energy values reported for the selected accelerators, and directly obtained results from executing models on the GPU, CPU, and TPU platforms to estimate the EPB and GOPS for each model and graph dataset.

We compared each hardware accelerator on the models supported by them, as outlined in their papers. For the models used in our analysis (Table 3), the GRIP and HyGCN accelerators support processing GCN, GraphSAGE, and GIN; EnG supports GCN and GraphSAGE; and HW_ACC supports GCN and GAT. The ReRAM-based hardware accelerators, ReGNN and ReGraphx, were used for the GCN and GraphSAGE model comparisons. As mentioned, one of the key attributes of *GHOST* is its versatility in accommodating a diverse set of GNN models, enabling it to support the different models used in our analysis.

*4.6.1 Throughput comparison*

Figure 10 shows the GOPS throughput comparison for *GHOST* with the computing platforms and GNN hardware accelerators considered. Our accelerator achieves on average 102.3×, 325.3×, 40.5×, 10.2×, 12.6×, 150.6×, 1699.0×, 1567.5×, 584.4× better GOPS when compared to GRIP, HyGCN, EnG, HW_ACC, ReGNN, ReGraphx, TPU, CPU, and GPU, respectively. While *GHOST* demonstrates promising improvements across all GNN models and datasets, the largest GOPS improvements are observed with the GIN graph dataset used for graph classification tasks. Across all datasets, processing GIN yielded on average 87.4× more GOPS when compared to the GNN hardware accelerators and 2168.9× when compared to GPU, CPU and TPU. This can be attributed to the small sizes of the graphs in each GIN dataset. These results are also consistent with those in Section 4.4, which illustrated that the partitioning optimization had the greatest impact on processing the graphs associated with the GIN model, leading to significant speedup. These findings highlight *GHOST*'s proficiency in handling diverse graph processing tasks. Also, there are significant GOPS improvements with *GHOST* for the GraphSAGE model, where it performed on average 100.9× better than other GNN accelerators and 1743.1× better than the GPU, CPU and TPU. This showcases how our accelerator is efficiently able to handle complex models, as it was able to overcome the complexity associated with supporting the sampling technique used in GraphSAGE.

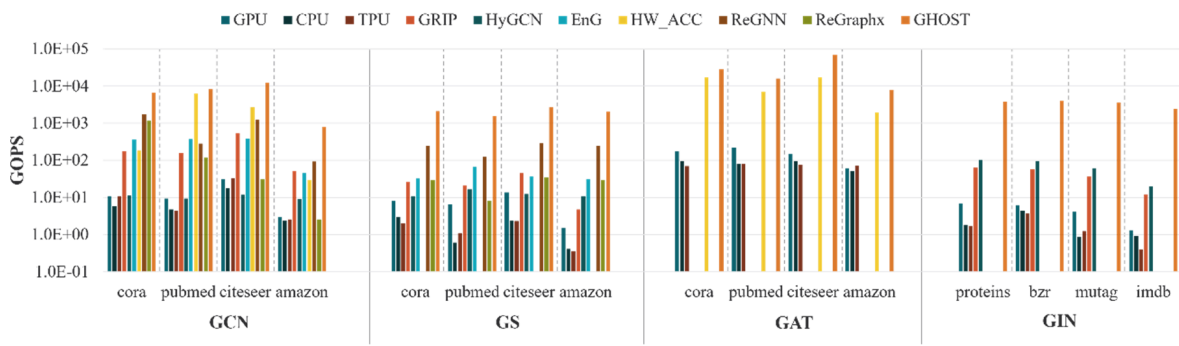

Figure 10: Throughput comparison between GPU, CPU, TPU, GNN hardware accelerators and *GHOST*.

*4.6.2 Energy Efficiency Comparison*

The EPB comparison results for *GHOST* with the computing platforms and GNN accelerators considered are shown in Figure 11. On average, *GHOST* attains 11.1×, 60.5×, 3.8×, 85.9, 15.7×, 313.7×, 24276.7, 6178.8×, 2585.3× lower EPB compared to GRIP, HyGCN, EnG, HW_ACC, ReGNN, ReGraphx, TPU, CPU and GPU, respectively. Our accelerator exhibits lower EPB values across all the GNN models and the graph datasets. In particular, GCN, which is the most widely used GNN model, achieved the lowest EPB values, with an average EPB reduction of 116.7× with *GHOST*, when compared to the GNN hardware accelerators and an average reduction of 7120.4×, in comparison to GPU, CPU and TPU. This is due to GCN's uniform operation pofile, which involves processes each vertex independently based on its immediate neighbors only, without considering other vertices in the graph. Overall, the improved energy efficiency can be explained in terms of *GHOST*'s significant low latency operation in the optical domain, as well as its relatively low power consumption of 18W compared to the other hardware accelerators and compute platforms.

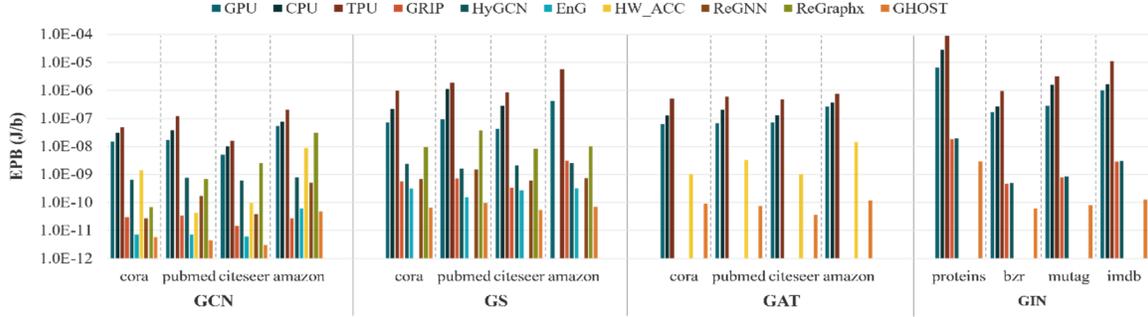

Figure 11: EPB comparison between GPU, CPU, TPU, GNN hardware accelerators and GHOST.

*4.6.3 EPB/GOPS Comparison*

The primary motivation for the design of *GHOST* is to obtain both an energy-efficient and high-throughput GNN acceleration. Accordingly, to showcase *GHOST*'s performance and energy improvements together, we show the EPB/GOPS comparison for *GHOST* against the computing platforms and GNN accelerators from prior work in Figure 12. It can be seen that *GHOST* achieves exceptionally lower EPB/GOPS values across all models and datasets. On average, *GHOST* attains 2.7e3×, 190.3e3×, 197.2×, 1.9e3×, 1.9e3×, 90.1e3×, 121.4e6×, 22.8e6×, 4.8e6× lower EPB/GOPS, compared to GRIP, HyGCN, EnG, HW_ACC, ReGNN, ReGraphx, TPU, CPU, and GPU, respectively. These results demonstrate how the cross-layer optimizations employed in *GHOST* across the circuit-level, device-level, and architecture-level along with performing most of the GNN operations in the optical domain, help to mitigate the various GNN inference challenges.

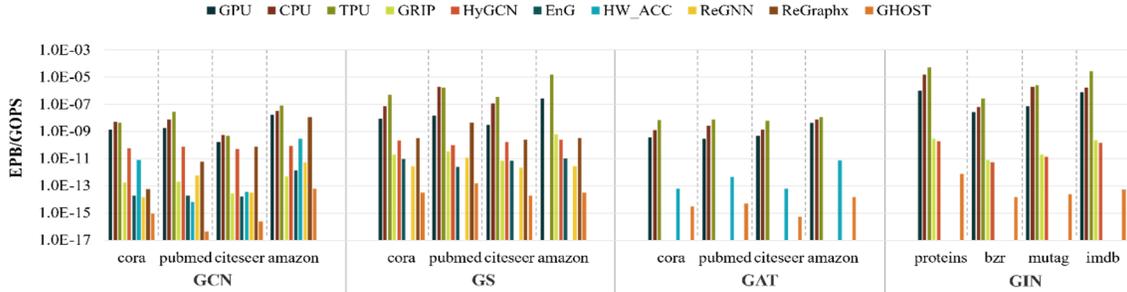

Figure 12: EPB/GOPS comparison between GPU, CPU, TPU, GNN hardware accelerators and *GHOST*.

## 5 CONCLUSION

In this paper, we presented the first silicon photonic GNN accelerator, called *GHOST*. In comparison to nine computing platforms and state-of-the-art GNN accelerators, our proposed accelerator exhibited throughput improvements of at least 10.2× and energy-efficiency improvements of at least 3.8×. These results demonstrate the promise of *GHOST* in terms of energy-efficiency and high-throughput inference acceleration for GNNs. This work focused on the hardware architecture design with silicon photonics and employed various device-, circuit-, and architecture-level optimizations. When combined with software optimization techniques to reduce the high memory requirements in GNNs, we expect that even better throughput and energy efficiency can be achieved. Furthermore, while various device and circuit design optimizations were employed in *GHOST,* several silicon photonic challenges can still be tackled to further improve throughput and energy efficiency. Such challenges include exploring alternate non-volatile optical memory cells [48] to minimize the needed digital buffering and opto-electric conversions. Another challenge hindering the progress of silicon photonics is the effect of fabrication process variations (FPVs) on the devices used [49]. Several techniques can be used to alleviate the impact of FPV, such as optical channel remapping, and intra-channel wavelength tuning [7].